\documentclass[aps,prl,twocolumn,floatfix]{revtex4-2}

\usepackage[english]{babel}
\usepackage[utf8]{inputenc}
\usepackage{amsthm}
\usepackage{mathtools}
\usepackage{physics}
\usepackage{xcolor}
\usepackage{graphicx}
\usepackage[left=23mm,right=13mm,top=35mm,columnsep=15pt]{geometry} 
\usepackage{adjustbox}
\usepackage{placeins}
\usepackage[T1]{fontenc}
\usepackage{lipsum}
\usepackage{csquotes}
\usepackage{hhline}
\usepackage{multirow}
\usepackage{dcolumn}  % Align table columns on decimal point
\usepackage{bm}  % bold math
\usepackage{lineno}

\usepackage{ulem}
\usepackage{cancel}

\usepackage[colorinlistoftodos, color=green!40, prependcaption]{todonotes}
\usepackage[pdftex, pdftitle={Article}, pdfauthor={Author}]{hyperref} % For hyperlinks in the PDF
\usepackage[nameinlink]{cleveref}
\newcommand{\code}[1]{{\lstinline!#1!}}
\newcommand{\NkT}{$Nk_{\mathrm{B}}T~$}
\newcommand{\kT}{$k_{\mathrm{B}}T~$}

\begin{document}

\title{On the Lattice Ground State of Densely Packed Hard Ellipses}

\author{S. Wagner$^1$, G. Kahl$^{1}$, R. Melnyk$^2$ and A. Baumketner$^{2}$}
\email[Correspondence email address: ]{andrij@icmp.lviv.ua}
\affiliation{$^1$Institute for Theoretical Physics, TU Wien, Wiedner Hauptstraße 8-10, A-1040 Wien, Austria,\\ $^2$Institute for Condensed Matter Physics, National Academy of Sciences of Ukraine, 1 Svientsistsky Str., Lviv, UA-79011, Ukraine}

\date{\today}

\begin{abstract}
Among lattice configurations of densely packed hard ellipses, Monte Carlo simulations are used to identify the so-called parallel and diagonal lattices as the two favourable states. The free energies of these two states are computed for several system sizes employing the Einstein Crystal method. An accurate calculation of the free energy difference between the two states reveals the parallel lattice as the state with lowest free energy. The origin of the entropic difference between the two states is further elucidated by assessing the roles of the translational and rotational degrees of freedom. 
\end{abstract}

\maketitle

%%% Letter
%% INTRODUCTION - ENTROPIC ORDERING IN HARD PARTICLE SYSTEMS (HISTORY AND CONCEPT)
Since the early 1950s, when computer simulations were first used to tackle problems of statistical mechanics numerically \cite{MetropolisTellerRosenbluth, Rosenbluth1954}, these methods have provided an extremely valuable insight into the pivotal role that  entropy plays in self-assembling systems. Notably, the long-standing notion of entropy being a thermodynamic potential driving any system into disorder was instantly challenged when simulations of hard spheres first indicated \cite{Alder1957, Wood1957a} (and later confirmed \cite{Pusey1986,Harland1997,Simeonova2004,Zaccarelli2009}) their crystallization, i.e., an ordering transition. Self-assembly was seen to occur into two alternative states, \textit{fcc} and \textit{hcp} lattices, with the question %over 
which of these two states possesses a lower free energy evolving into a  decade-long scientific debate - discussed vigorously in many theoretical studies using a variety of numerical and analytical methods  \cite{Radun2005, Woodcock1997, WoodcockRep}- before being settled in  favor of \textit{fcc} lattice \cite{NoyaAlmarza2015, BruceWilding1997, BolhuisMau1997, MauHuse1997, MiguelMarguta2007}.

In systems of hard ellipses the situation becomes even more involved, as in addition to the translational degrees of freedom (DOFs) also a rotational  DOF comes into play. Onsager was the first to point out the interplay between these two types of DOFs in the system of hard  rods~\cite{Onsager1949}, for which rotations were seen to play an increasingly important role as the spatial elongation of the particles increased, leading to a transition into the nematic phase, i.e. a partially ordered phase in which the orientations of the particles are all aligned while the respective positions remain disordered. Nematic phases are also the focal point of many theoretical studies of hard-core ellipses~\cite{VieillardBaron1972, FrenkelEppenga1985, CuestaFrenkel1990, Maeda2003}.
However, considerably less attention has been devoted to the crystallization transition in this system that occurs  upon increasing the density and results in the formation of a densely packed ordered state. In contrast to their shape-isotropic counterparts of hard disks, ellipses are able to arrange into a multitude of different lattice states, some of which were considered earlier \cite{BautistaOdriozola2014,CuestaFrenkel1990}.
However, the nature of these different lattice conformations is not yet understood. In  particular, the questions of how many stable structures can be expected and which of these is the most stable one have not been addressed so far.

The present contribution is dedicated to answering  exactly these  questions. Among all possible ellipse lattice states related to hexagonal order \cite{VieillardBaron1972}, we identify - by means of Monte Carlo (MC) simulations - two stable candidates: the so-called diagonal and the parallel lattice states, each of which is characterized by the particles' relative positions to one another as well as by their orientations with respect to the lattice direction. We then proceed to evaluate the free energy difference between the identified states by  explicit free energy methods, similarly to the approaches employed for the hard sphere crystal problem \cite{NoyaAlmarza2015}. Finally, one lattice is identified with the lowest free energy - or equivalently with the highest entropy - which is analyzed in terms of the particles' translational and orientational DOFs contributing to the lattice's thermodynamic stability. 

%%%%%%%%%%%%%%%%%%%
%%     MODEL     %%
%%%%%%%%%%%%%%%%%%%

\begin{figure}
    \centering
    \includegraphics[width=\linewidth]{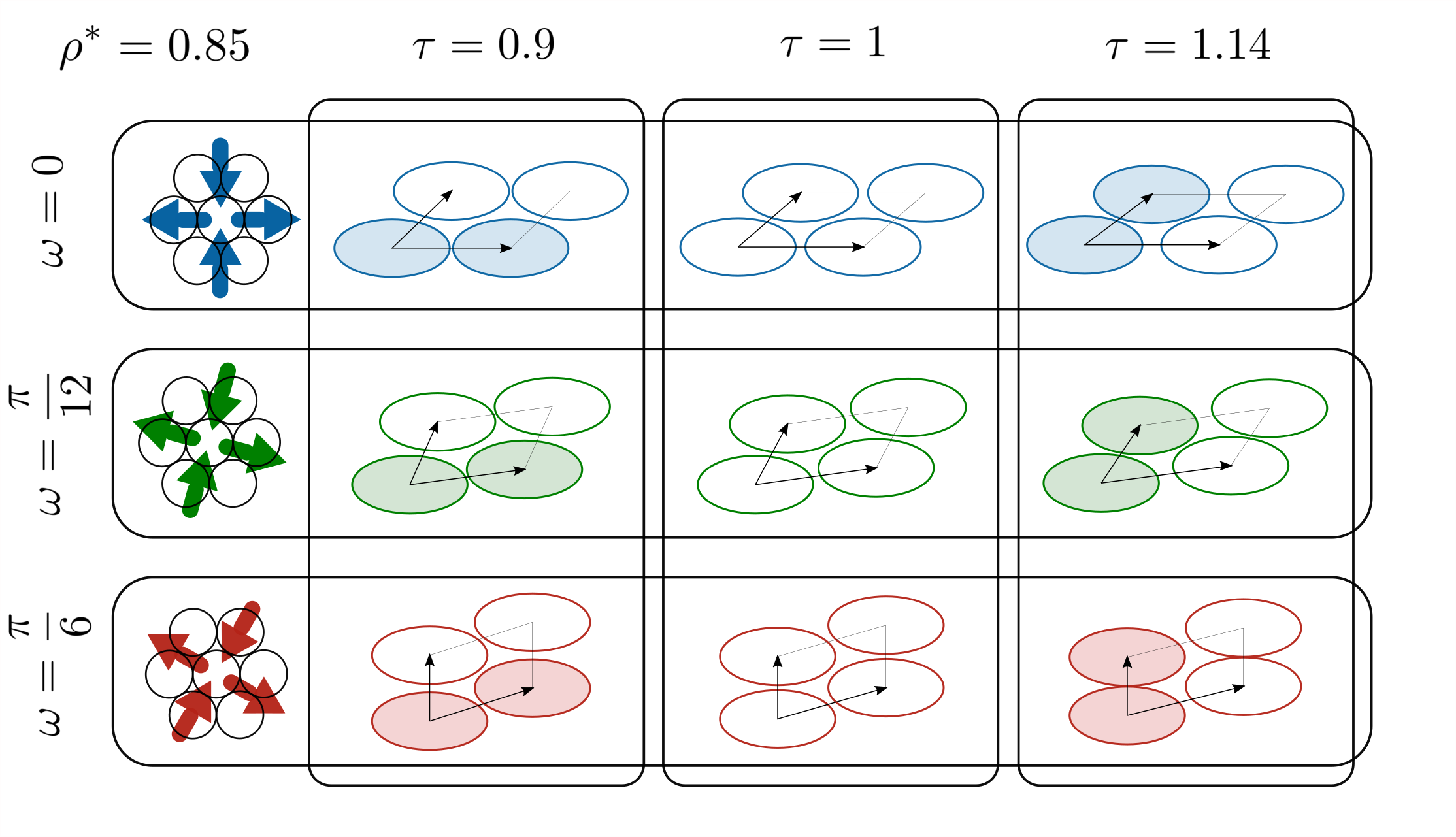}
    \caption{
    Transformation of a close packed lattice of disks (left column) into hexagonal-like lattices of ellipses. Each panel shows the unit cell of the lattice, as defined by lattice parameter $\omega$ (rows) and $\tau$ (columns), and spanned by the unit cell vectors; see text. The colored particles mark a pair of ellipses that is moved closer to one another via the $\tau$-transformation.
    }
    \label{fig:param}
\end{figure}

We consider crystalline, high-density states of hard elliptic particles of aspect ratio $\kappa=2$. We name the described lattice states \textit{hexagonal-like} lattices as these structures can be derived via deforming a hexagonal lattice of disks. In these hexagonal-like lattices all ellipses are oriented in the same direction having six nearest neighbours each. The transformation of a lattice of disks into a lattice of ellipses was first described by Vieillard-Baron \cite{VieillardBaron1972} at close-packed density $\rho_{\mathrm{cp}}$. Here we generalize this approach to arbitrary densities $\rho^*$ (defining $\rho^{\ast}=\frac{\rho}{\rho_{\mathrm{cp}}}$) by performing two subsequent transformations, each of them being characterized by a parameter, which allows us to parameterize the resulting lattice states as seen in \Cref{fig:param}. 
First, we apply an affine transformation, $\mathbf{H_1}$ that deforms disks into ellipses of aspect ratio $\kappa$, see \Cref{equ:01}. The angle $\omega$ at which this transformation  is applied to the hexagonal lattice of disks can be chosen arbitrarily with $\omega\in[0,\pi/6)$ as the lattice is characterized by a dihedral symmetry $\mathcal{D}_6$. We introduce the angle $\omega$ related to the rotation $\mathbf{R}(\omega)$ as the first lattice parameter, defining the transformation:
\begin{equation}\label{equ:01}
    \mathbf{H_1}(\kappa)\cdot\mathbf{R}(\omega)=\Big(\begin{smallmatrix}\sqrt{\kappa} & 0 \\ 0 & \frac{1}{\sqrt{\kappa}}\end{smallmatrix}\Big)\cdot
    \Big(\begin{smallmatrix}\cos{\omega} & -\sin{\omega} \\ \sin{\omega} & \cos{\omega}\end{smallmatrix}\Big)\text{.}
\end{equation}
Upon expanding the close-packed lattice state to lower densities $\rho^{\ast}<1$, a second lattice parameter, $\tau$, is introduced as the ratio between the expansion rates along the two spatial directions, visualized in \Cref{fig:param}. The transformation capturing the non-uniform expansion along the two lattice directions is given by:
\begin{equation}
    \mathbf{H_2}(\rho^{\ast},\tau)=\frac{1}{\sqrt{\rho^{\ast}}}\Big(\begin{smallmatrix}\sqrt{\tau} & 0 \\ 0 & \frac{1}{\sqrt{\tau}}\end{smallmatrix}\Big) .
\end{equation}

%%%%%%%%%%%%%%%%%%%%%%%%%%%
%% LATTICE OPTIMIZATION  %%
%%%%%%%%%%%%%%%%%%%%%%%%%%%

\begin{figure}
    \centering
    \includegraphics[width=0.75\linewidth]{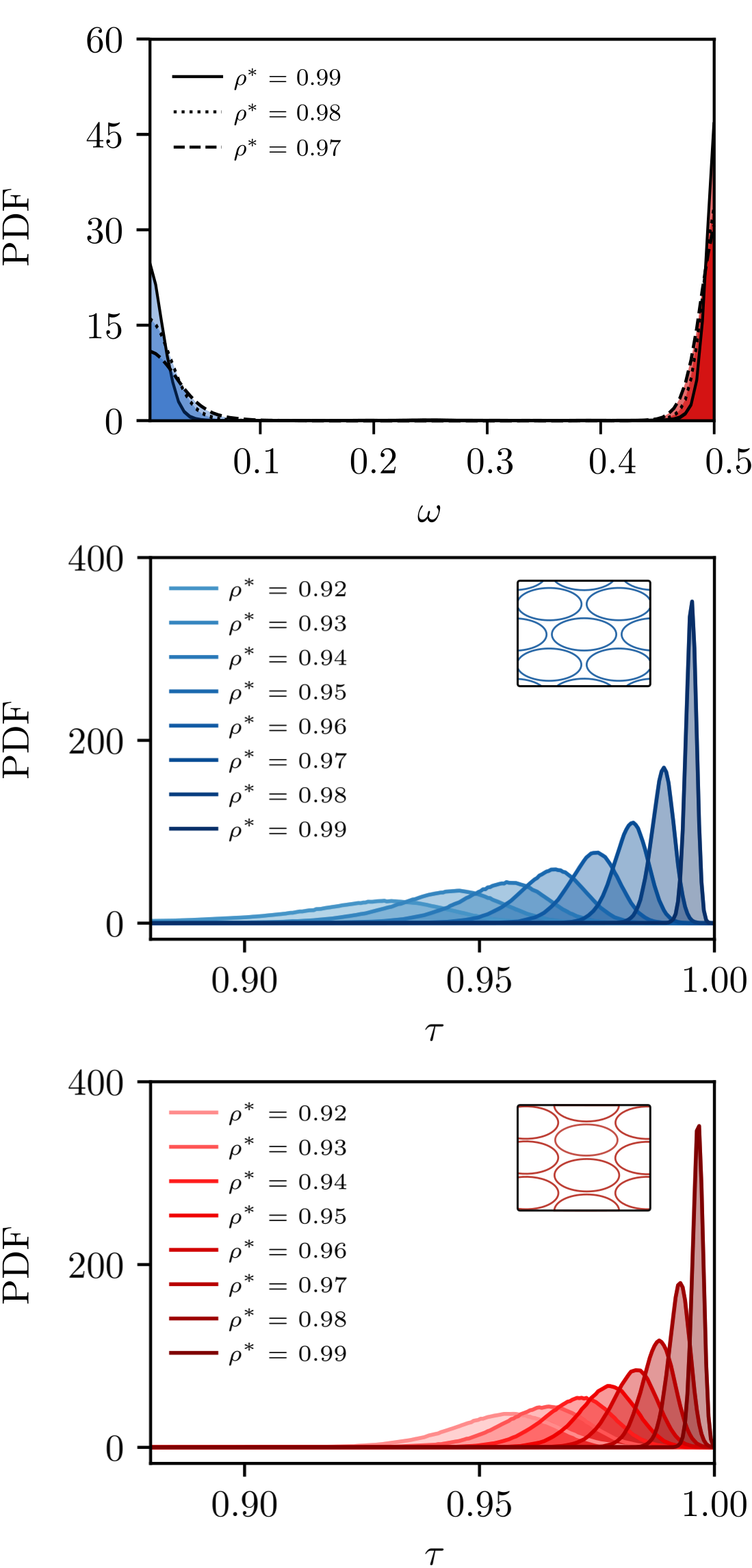}
    \caption{
    PDFs as obtained from lattice MC simulations for determining the optimal values $\omega_{\mathrm{opt}}$ and $\tau_{\mathrm{opt}}$ for various densities $\rho^{\ast}$ (as labeled). Top panel: PDFs as functions of $\omega$ from simulations with $\omega$-moves at fixed $\tau=1$. We find two distinct states, the diagonal ($\omega^{\mathrm{opt}}=0$ in blue) and the parallel state ($\omega_{\mathrm{opt}}=\frac{\pi}{6}$ in red). Central and bottom panels: PDFs as functions of $\tau$ from simulations with $\tau$-moves at fixed $\omega_{\mathrm{opt}}=0$ (central) and $\omega_{\mathrm{opt}}=\frac{\pi}{6}$ (bottom). Optimal values $\tau_{\mathrm{opt}}$ are identified as maxima of the respective PDF.
    }
    \label{fig:TAUS}
\end{figure}
We now are interested in the following question: which state, defined by $(\omega_{\mathrm{opt}}, \tau_{\mathrm{opt}})$, is entropically the most favorable one? To answer this question,  we apply a specifically designed lattice MC method \cite{BruceWilding1997}, that probes different lattice states by proposing lattice-parameter MC-moves, $\Delta \omega$ and $\Delta \tau$, with a rate of 5\% during he course of a single particle move MC-simulation that involves particles' displacements and rotations. In a first step, we implement lattice moves for $\omega$ at fixed $\tau=1$. In an ensemble of $N=100$ particles, we perform a total of $60$ parallel runs, with initial lattice parameters $\omega_{\mathrm{init}}$ evenly distributed within the range $[0,\pi/6)$. The resulting probability density functions (PDFs) are shown in the top panel of \Cref{fig:TAUS}. We see that stability occurs at two highest symmetry states, i.e. for $\omega_{\mathrm{opt}}=0$ and $\omega_{\mathrm{opt}}=\frac{\pi}{6}$, which  we term  the \textit{diagonal} and \textit{parallel} lattice, respectively. In contrast, probabilities for most of the other $\omega$-values vanish. With this clear indication of two very distinct optimal lattice states, the second optimization step with respect to the parameter $\tau$ is applied, focusing on the two optimal $\omega$-parameters only. This second optimization step leads to PDFs shown in the central and the bottom panel of \Cref{fig:TAUS}, considering the diagonal and parallel lattice states both in rectangular box geometries and at various densities.

It is seen that as the density is increased, the distributions become more sharply peaked around certain optimal values. This trend is consistent between simulations with constant $\tau$ or $\omega$. Specifically, lattices were constructed by replicating the unit cells of ellipses (as shown in \Cref{fig:param}) along the directions of the unit cell vectors, thereby creating skewed simulation boxes  defined by box side lengths $L_{\mathrm{x}}$ and $L_{\mathrm{y}}$, as well as an enclosing angle $\alpha$. Instead of optimizing the lattice state by varying the lattice parameters $\omega$ and $\tau$, we now let the simulation box geometry adjust according to the ellipses' most favourable arrangement. We introduce MC-moves acting solely on the box parameters - $L_{\mathrm{x}}$, $L_{\mathrm{y}}$ and $\alpha$ - while keeping the volume constant \cite{Baumketner2022}. Our simulations uncover two dominant lattice configurations at high densities: a) those defined by $\alpha_{\textrm{min}}$ that correspond to the diagonal state and b) those defined by $\alpha_{\textrm{max}}$ corresponding to the parallel state. It can be shown~\cite{Baumketner2022} that $ \alpha_{\textrm{min}}=\cos^{- 1}\sqrt{\frac{3}{3 + \kappa^{2}}}$ and $\alpha_{\textrm{max}}=\frac{\pi}{2} - \cos^{-1}\sqrt{\frac{3\kappa^{2}}{3 + \kappa^{2}}}$, where $\kappa$ is the aspect ratio.

\begin{figure}
    \centering
    \includegraphics[width=0.9\linewidth]{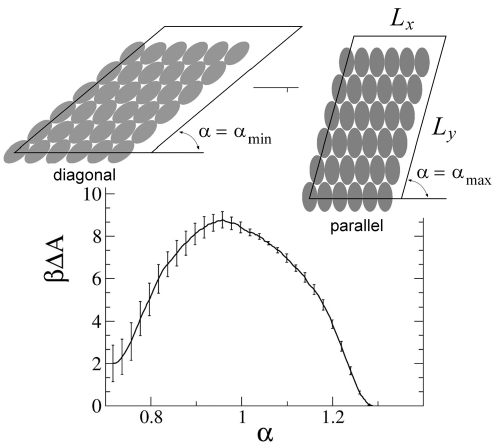}
    \caption{
    Free energy profile $\Delta A(\alpha)=A(\alpha)-A(\alpha_{\rm max})$
    shifted to zero at $\alpha=\alpha_{\rm max}$
    as obtained for the system composed of $N$=36 particles at $\rho^{*}=0.98$ in umbrella sampling simulations. Error bars were estimated as standard deviations over 10 independent trajectories.}
    \label{fig:reweight}
\end{figure}

In a subsequent step, we used the umbrella sampling method~\cite{umbrella} to determine the free energy difference between the diagonal and the parallel state. We applied a biasing potential on $\alpha$ in order to force the lattice to find its equilibrium configuration at the pinned value of the box angle. Multiple bins were considered so as to construct a path connecting the two states. Simulations were performed for each bin and then  combined in the  re-weighting procedure~\cite{rev} to produce the free energy profile as a function of the box angle $\alpha$, shown in \Cref{fig:reweight} as $\Delta A(\alpha) = A(\alpha)-A(\alpha_{\textrm{max}})$. Its appearance, featuring two minima and one maximum confirms that there are two stable states. Furthermore, the parallel state appears as the global free energy minimum while the diagonal state corresponds to a local minimum. We find the free energy difference (FD) between the two states, $\Delta A_{\textrm{d}-\textrm{p}}=\Delta A(\alpha_{\textrm{min}})$ to be 2.0$\pm$0.9 \kT.

%%%%%%%%%%%%%%%%%%%
%% FREE ENERGIES %%
%%%%%%%%%%%%%%%%%%%
This value for the FD is small but statistically significant. Still, the FD may result from a systematic error stemming from the finite size of the simulated system ($N$=36) rather than representing the true free energy difference between the two states. In an effort to determine which of the two scenarios takes place - i.e., either finite-size effects or a real FD sustained in the thermodynamic limit - we performed a finite-size scaling analysis. To that end, $\Delta A_{\textrm{d}-\textrm{p}}$ was computed for  several system sizes $N$, with the aim of performing an extrapolation towards the limit of $N \to \infty$. As the umbrella sampling simulations become cost-prohibitive for large $N$, we instead used the Einstein crystal method~\cite{FrenkelLadd1984}, more specifically an implementation proposed by Vega~et~al~\cite{NoyaVega2007,VegaNoya2008}. In this method, the total free energy $A$ of the target state, defined by Hamiltonian $\mathcal{H}$, is computed by connecting it via a thermodynamic path to a reference state of known free energy. This reference state is an Einstein crystal (EC), i.e. a lattice of independent, classical harmonic oscillators defined by an Hamiltonian $\mathcal{H}_{\mathrm{EC}}$ and a free energy $A_0$. The EC is linked to the target lattice state via an intermediate mixed state defined by the Hamiltonian $\mathcal{H}_{\lambda}=\mathcal{H}+(1-\lambda)\mathcal{H}_{\mathrm{EC}}$, with $\lambda$ as a tuning parameter. The free energy change $\Delta A_1$ associated with passing from the EC to the mixed state $\mathcal{H}_{\lambda}$ with $\lambda=0$ is computed using the umbrella sampling method \cite{Frenkel2023}. In a second step, the harmonic potential is switched off (via $\lambda \to 1$), thus  taking the system to the original Hamiltonian and completing the integration path. The associated free energy cost of the last step is expressed via the following integral:
\begin{align}
    \Delta A_2=-\int_{\lambda=0}^{1}\Big\langle \sum_{i=1}^{N}(\Lambda_{\rm E}^{\rm t}\Delta \mathbf{r}_i^2 + \Lambda_{\rm E}^{\rm r}\Delta \phi_i^2)\Big\rangle_{\lambda}d\lambda \text{ ,}
    \label{A2}
\end{align}
where $\Delta \mathbf{r}_{i}=\mathbf{r}_i-\mathbf{r}_{i,0}$ and $\Delta \phi_i=\phi_i-\phi_{i,0}$ describe the translational and orientational displacements of particle $i$ with respect to the defined equilibrium lattice states $\mathbf{r}_{i,0}$ and $\phi_{i,0}$. The constants $\Lambda_{\textrm{E}}^{\textrm{t}}$ and $\Lambda_{\textrm{E}}^{\textrm{r}}$ are adjustable parameters that set the strength of the harmonic force in the EC. Finally, the total free energy is expressed  as the sum of three terms $A$ = $A_0$ + $\Delta A_1$ + $\Delta A_2$. 

Since the EC is the same reference state for both the parallel and diagonal lattice, the term $A_0$ drops from the difference $\Delta A_{\mathrm{d-p}}$. Further, the contribution $\Delta A_1$ can be made numerically very small by an appropriate choice of the constants $\Lambda_{\textrm{E}}^{\textrm{t}}$ and $\Lambda_{\textrm{E}}^{\textrm{r}}$. In our computations, this leaves the integral $\Delta A_2$ as the only non-zero contribution to the free energy difference $\Delta A_{\mathrm{d-p}}$. We compute this difference for boxes of two different geometries, a rectangular and a skewed simulation box, at a density $\rho^{\ast}=0.98$. For the rectangular box we construct lattices using the optimal lattice parameters $\omega_{\mathrm{opt}}$ and $\tau_{\mathrm{opt}}$. The considered systems varied in size from $N=32$ to $N=398$ particles. The number of particles in each lattice was chosen such that a box aspect ratio of $\sim 1$ is achieved, making the boxes almost square-shaped. For the skewed boxes the optimal geometry  was extracted from simulations with fixed box angles $\alpha$ at $\alpha_{\textrm{min}}$ and $\alpha_{\textrm{max}}$. The number of particles was chosen as $N=n^2$, where $n$ varies from 6 to 18. 

\begin{figure}
    \includegraphics{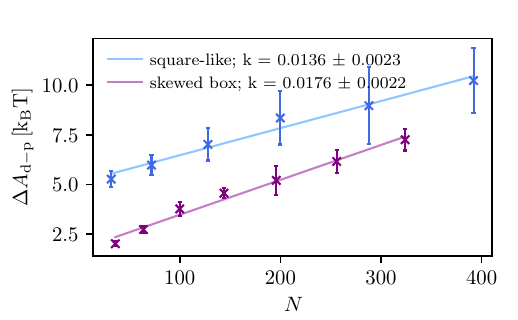}
    \caption{\label{fig:deltaA_linreg} FD between the diagonal (d) and parallel (p) lattices at $\rho^{\ast}=0.98$ for different system sizes $N$ and in two box geometries. At finite sizes the box geometry has a significant effect on the FD. When applying linear regression models to obtain the thermodynamic limit, given by constant $k$ results in both box geometries agree within error bars. }
\end{figure}

We evaluate $\Delta A_2$ in \Cref{A2} numerically with the results shown in \Cref{fig:deltaA_linreg}. For each box type we perform a finite-size analysis by fitting $\beta \Delta A_{\mathrm{d-p}}(N)$ to $f(N)=k N^{b-1}+c$, where constants $k$, $b$, and $c$ are treated as adjustable parameters. For both boxes the constant $b$ was found to be close to 2, suggesting that the linear function $f(N)=k N+c$ is able to represent the simulation data best. We therefore proceed with a linear regression model. In the thermodynamic limit $N\rightarrow\infty$, the FD per particle $\Delta A^{\ast}_{\mathrm{d-p}}=N^{-1} \Delta A_{\mathrm{d}-\mathrm{p}}$ equates to the constant $k$. The linear regression applied to our simulated data as shown in \Cref{fig:deltaA_linreg} yields $\Delta A^{\ast}_{\mathrm{d}-\mathrm{p}}=(0.0136 \pm 0.0023)$\kT for the square box and $\Delta A^{\ast}_{\mathrm{d-p}}=(0.0176 \pm 0.0022)$\kT for the skewed box with a confidence of $90\%$. Confirming the high credibility of the reported results, the estimates are in agreement within error bars, strongly suggesting that the FD between the diagonal and parallel state is real and not an artefact of the employed simulation protocol.

Next, we focus on the microscopic origin of this FD. \Cref{A2} provides an excellent tool for this task as it enables a clear separation of the total free energy into contributions originating from the translational and the orientational DOFs, i.e.  $\Delta A^{\mathrm{t}}_2 = - \int\langle\sum_i^N\Lambda^{\rm t}_{\rm E} \Delta\mathbf{r}_i^2\rangle_{_{\lambda}} d\lambda$ and $\Delta A^{\mathrm{r}}_2= - \int\langle\sum_i^N\Lambda^{\rm r}_{\rm E} \Delta\phi_i^2\rangle_{_{\lambda}} d\lambda$, respectively. This separation will facilitate a mechanistic understanding of the origin of the FD. 
It is important  to point out that there is no strict distinction between the translational and rotational contributions to the FD as the DOFs are coupled. Nevertheless,
$\Delta A^{\mathrm{t}}_2$ and $\Delta A^{\mathrm{r}}_2$ can be seen as a useful measure of the phase-space volume roughly accessible to the respective DOFs. We also note that these integrals depend on the actual values of $\Lambda^{\textrm{t}}_{\textrm{E}}$ and $\Lambda^{\textrm{r}}_{\textrm{E}}$; still, we find that this dependence is weak and does not affect the conclusions at the qualitative level.
As an example, we report that $\Delta A_2^{\mathrm{t,d-p}}=\Delta A_2^{\mathrm{t},\textrm{d}}-\Delta A_2^{\textrm{t},\textrm{p}} =(0.260 \pm 0.041$) \NkT for the translational DOFs and
$\Delta A_2^{\mathrm{r,d-p}}=\Delta A_2^{\textrm{r},\textrm{d}}-\Delta A_2^{\textrm{r},\textrm{p}} =(-0.235 \pm 0.022$) \NkT for the rotational DOF for the system with $N=392$ particles and a particular set of $\Lambda_{\rm E}$'s. Remarkably, the contributions come with a differing sign, suggesting that the translational DOFs enhance the stability of the parallel state while the rotational DOF seem to destabilize it.

We performed further tests to elucidate the role of different DOFs. The FD for a system of $N=36$ particles in a skewed simulation box is $\Delta A_{\textrm{d}-\textrm{p}}=(2.0\pm0.1)$\kT. To probe the effect of particle rotations on this value, we repeated the computations for the same system but allowing only particle translations, i.e. freezing rotational DOFs. Here we obtained a FD $\Delta A_{\textrm{d}-\textrm{p}}=(5.6\pm0.1)$\kT, indicating that the parallel state does offer a much greater translational freedom to the ellipses. In a similar manner, when we freeze the translational DOFs, the resulting FD becomes a non-negative value $\Delta A_{\textrm{d}-\textrm{p}}=(0.6\pm0.1)$\kT, opposed to the analysis based on \Cref{A2}. Interestingly, in the system with frozen translational DOFs the parallel lattice turns out to offer a larger free volume accessible through rotations, when in the system with all DOFs active it is the diagonal lattice. This result highlights the impact of the coupling of the rotational to the translational DOFs as it drastically raises the rotational entropy within the diagonal lattice. Overall, however, the higher rotational entropy in the diagonal lattice cannot outweigh the advantage achieved by the translational DOFs within the parallel lattice.

The importance of the rotational DOF is further highlighted when we again compute FD in a system of ellipses whose orientational DOFs are frozen, but now in a simulation box that has relaxed to the equilibrium geometry specific to the system with frozen orientational DOFs. Here we found that the FD among all states, including the parallel and diagonal one, vanishes completely.

Taken together, these results show that ellipses require both translations and rotations in order to develop a preferred thermodynamic state at high densities. Further, they elevate the role of coupling of rotational and translational DOFs in letting the parallel lattice state develop this thermodynamic preference.

%%% Bibliography

\end{document}